\documentclass[hyper]{JHEP3}
\usepackage{latexsym,epsf}
\usepackage{graphicx}
\usepackage{bm}
\usepackage{psfig,epsfig}

\newcommand{\za}{\alpha}

\relax

\newcommand{\be}{\begin{equation}}
\newcommand{\ee}{\end{equation}}
\newcommand{\bea}{\begin{eqnarray}}
\newcommand{\eea}{\end{eqnarray}}
\newcommand{\ba}{\begin{array}}
\newcommand{\ea}{\end{array}}
\newcommand{\bc}{\begin{center}}
\newcommand{\ec}{\end{center}}

\newcommand{\bds}{\begin{description}}
\newcommand{\eds}{\end{description}}

\newcommand{\bsg}{$b\rightarrow s + \g$}

\newcommand{\atal}{{\it et al.}}

\def\321{$SU(3)\times SU(2)\times U(1)$}

\def\b{\beta}
\def\bsg{$ b \rightarrow s + \g$}
\def\g{\gamma}

\def\d{\delta}

\def\f{\frac}

\def\l{\lambda}
\def\n{\nu}
\def\m{\mu}

\def\ssc{\scriptscriptstyle}
\def\wtl{\widetilde}

\def\vckm{V_{\!\mbox{\tiny CKM}}}
\title{Quark Loop Contributions  to Neutron EDM 
from R-parity Violation}

\author{Chan--Chi Chiou, Otto C. W. Kong, and Rishikesh
D. Vaidya\\
Department of Physics, National Central University,
Chung-Li, TAIWAN 32054.}

\abstract{
We present a detailed analysis together with numerical calculations on
one-loop contributions to the neutron electric dipole moment from
supersymmetry without R parity, focusing on the quark-scalar loop
contributions. Complete formulas are given for the various
contributions through the quark dipole operators.  Analytical expressions 
illustrating the explicit role of the R-parity violating parameters
are given following perturbative diagonalization of mass-squared
matrices for the scalars. Dominant contributions come from the combinations
$B_i^{\ast} \l^{\prime}_{ij1}$ for which we obtain robust
bounds. Even if the R-parity violating couplings are real, CKM
phase does induce RPV contribution and for some cases such a contribution 
is as strong as contribution from phases in the R-parity violating couplings.  
Hence, we have bounds directly on $|B_i^{\ast} \l^{\prime}_{ij1}|$
even if the parameters are all real.
}
\preprint{NCU-HEP-k022}
\keywords{neutron EDM, R-parity Violation, Supersymmetry}
\begin{document}
\section{Introduction} 
Forty years after the discovery of CP-violation\cite{cpv}, its
experimentally observed effects in the K and B-meson systems
\cite{KBcpv} are generally compatible with the standard model (SM)
predictions with the Kobayashi-Maskawa (KM) phase as its sole source. Electric dipole
moments (EDMs) of elementary particles are likely to be the best candidate for the
evidence of CP violating physics beyond the SM. The presence of an EDM itself violates
CP\cite{Landaucpv}. Neutron being electrically neutral, is studied extensively by 
experimentalists, for a possible nonvanshing EDM. In the SM, the KM phase generated EDM of 
neutron is of the order $10^{-32}$ ecm \cite{nanopoulousEDM}, being still way below the
current experimental bound of $6.3\times10^{-26}\,\mbox{ecm}$
\cite{expNEDM}.

Many extensions of the SM are expected to give potentially large EDM contributions. 
For instance, in the minimal supersymmetric standard model (MSSM), which is no
doubt the most popular candidate theory for physics beyond SM, the plausible 
extra EDM contribution of substantial magnitude has led to the SUSY-CP problem\cite{SUSYCP}. 
If one simply takes the minimal supersymmetric spectrum of the SM and imposes
nothing more than the gauge symmetries while admitting soft SUSY breaking, 
the generic supersymmetric standard model (GSSM)\cite{GSSM} would result.
When the large number of baryon or lepton number violating terms are
removed by imposing an {\it ad hoc} discrete symmetry called R-parity, one obtains
the MSSM Lagrangian. GSSM is a complete theory of  supersymmetry (SUSY)
without R-parity, where all kinds of R-parity violating (RPV) terms are
admitted without bias. It is  generally better motivated than {\it ad hoc} versions
of RPV theories. The MSSM itself, without extension such as adding SM
singlet superfields and admitting violation of lepton number, cannot accommodate
neutrino mass mixings and hence oscillations \cite{neutrino}. Given SUSY, the GSSM is
actually conceptually the simplest framework to accommodate the
latter. The large number of {\it a priori} arbitrary R-parity violating terms do make 
phenomenology complicated. However, the origin of the (pattern of) values for the
couplings may be considered to be on the same footing as that of the SM Yukawa
couplings. For example, it has been shown in \cite{u1} that one can understand the origin, 
pattern and magnitude of all the R-parity violating terms as a result of a 
spontaneously broken anomalous Abelian family symmetry model.

The simplest approach to study quantum corrections to the neutron EDM
is to analyze loop contributions to the $u$ and $d$ quark dipole
moments and obtain the overall neutron number based on the valence
quark model\cite{HKP}. The neutron EDM is given by
\be \label{vqm}
d_n = {1\over 3} \left (4\, d_d -d_u \right) \, \eta \;,
\ee
where $\eta \simeq 1.53$ is a QCD correction factor from
renormalization group evolution\cite{IN1,QCD}. In the case that only  trilinear 
RPV couplings are admitted, it  has been  shown that nonvanishing results  come 
in only starting at two-loop level\cite{2loop}. Striking one-loop contributions are, 
however, identified and discussed based on the GSSM framework\cite{nedm,chun}.
Similarly, flavor off-diagonal dipole moment contributing to 
the case of  the $b\rightarrow s \g$ in \cite{bsg} decay and that of 
the $\m \rightarrow e \g$ \cite{mueg} decay at one-loop level have also been presented.  
That is the approach taken here. In Ref.\cite{nedm}, the one loop EDM contributions from 
the gluino loop, chargino-like loop, neutralino-like loop are studied numerically in
some details. It shows that the RPV parameter combination
$\mu_i^*\lambda^{\!\prime}_{i\scriptscriptstyle 1\!1}$ dominates with
small sensitivity to the value of $\tan\!\beta$.  The experimental
bound on neutron EDM is used to constrain the model parameter space,
especially the RPV part.  The deficiency of Ref.\cite{nedm}, missing
details of quark(-scalar loop), is made up here.  We give complete
1-loop formulas for the type of contributions to the EDMs of the up-
and down-sector quarks, with full incorporation of family mixings. We
present numerical analysis of quark-scalar loop contributions from all
possible combinations of RPV parameters. Besides the familiar
$\mu_i^*\lambda^{\!\prime}_{i\scriptscriptstyle 1\!1}$, there are a
list of combinations of the type $B_i^*\lambda^{\!\prime}_{ijk}$ which
are particularly interesting and will be the focus of this paper. Such
a calculation has not been reported before. Hence, we obtain bounds on
combinations of RPV couplings which are otherwise unavailable. Our
study also guides plausible experimental searches for RPV SUSY signals
at the next generation neutron EDM experiments.

We skip the details of the MSSM contribution to neutron EDM since it
has already been extensively discussed in the literature \cite{nedm-mssm}.

\section{Formulation and Notation} 
We summarize the model here while setting the notation. Details of
the formulation adopted is elaborated in Ref.\cite{GSSM}.
The most general renormalizable superpotential for the supersymmetric
SM (without R-parity) can be written  as
\small\begin{eqnarray}
W \!\! &=& \!\varepsilon_{ab}\Big[ \mu_{\alpha}  \hat{H}_u^a \hat{L}_{\alpha}^b
+ h_{ik}^u \hat{Q}_i^a   \hat{H}_{u}^b \hat{U}_k^{\scriptscriptstyle C}
+ \lambda_{\alpha jk}^{\!\prime}  \hat{L}_{\alpha}^a \hat{Q}_j^b
\hat{D}_k^{\scriptscriptstyle C}
\nonumber \\
&+&
\frac{1}{2}\, \lambda_{\alpha \beta k}  \hat{L}_{\alpha}^a
 \hat{L}_{\beta}^b \hat{E}_k^{\scriptscriptstyle C} \Big] +
\frac{1}{2}\, \lambda_{ijk}^{\!\prime\prime}
\hat{U}_i^{\scriptscriptstyle C} \hat{D}_j^{\scriptscriptstyle C}
\hat{D}_k^{\scriptscriptstyle C}  \; ,
\end{eqnarray}\normalsize
where  $(a,b)$ are $SU(2)$ indices, $(i,j,k)$ are the usual family (flavor)
indices, and $(\alpha, \beta)$ are extended flavor index going from $0$ to $3$.
In the limit where $\lambda_{ijk}, \lambda^{\!\prime}_{ijk},
\lambda^{\!\prime\prime}_{ijk}$ and $\mu_{i}$  all vanish,
one recovers the expression for the R-parity preserving case,
with $\hat{L}_{0}$ identified as $\hat{H}_d$. Without R-parity imposed,
the latter is not {\it a priori} distinguishable from the $\hat{L}_{i}$'s.
Note that $\lambda$ is antisymmetric in the first two indices, as
required by  the $SU(2)$  product rules, as shown explicitly here with
$\varepsilon_{\scriptscriptstyle 12} =-\varepsilon_{\scriptscriptstyle 21}=1$.
Similarly, $\lambda^{\!\prime\prime}$ is antisymmetric in the last two
indices, from $SU(3)_{\scriptscriptstyle C}$.

The large number of new parameters involved, however, makes the theory difficult to 
analyze.  An optimal parametrization, called the single-VEV parametrization (SVP) has 
been advocated\cite{ru2} to make the the task manageable. Here, the choice of an optimal 
parametrization mainly concerns the 4 $\hat{L}_\alpha$ flavors. Under the SVP,
flavor bases are chosen such that :
1) among the $\hat{L}_\alpha$'s, only  $\hat{L}_0$, bears a VEV,
{\it i.e.} {\small $\langle \hat{L}_i \rangle \equiv 0$};
2)  {\small $h^{e}_{jk} (\equiv \lambda_{0jk})
=\frac{\sqrt{2}}{v_{\scriptscriptstyle 0}} \,{\rm diag}
\{m_{\scriptscriptstyle 1},
m_{\scriptscriptstyle 2},m_{\scriptscriptstyle 3}\}$};
3) {\small $h^{d}_{jk} (\equiv \lambda^{\!\prime}_{0jk} =-\lambda_{j0k})
= \frac{\sqrt{2}}{v_{\scriptscriptstyle 0}}{\rm diag}\{m_d,m_s,m_b\}$};
4) {\small $h^{u}_{ik}=\frac{\sqrt{2}}{v_{\scriptscriptstyle u}}
V_{\!\mbox{\tiny CKM}}^{\!\scriptscriptstyle T}\;
{\rm diag}\{m_u,m_c,m_t\}$}, where
${v_{\scriptscriptstyle 0}} \equiv  \sqrt{2}\,\langle \hat{L}_0 \rangle$
and ${v_{\scriptscriptstyle u} } \equiv \sqrt{2}\,
\langle \hat{H}_{u} \rangle$.
The big advantage of here is that the (tree-level) mass
matrices for {\it all} the fermions  {\it do not} involve any of
the trilinear RPV couplings, though the approach makes {\it no assumption}
on any RPV coupling including even those from soft SUSY breaking;
and all the parameters used are uniquely defined, with the exception of
some possibly removable phases. 

The soft SUSY breaking part of the Lagrangian in GSSM can be written as follows : 
\begin{eqnarray}
V_{\rm soft} &=& \epsilon_{\!\scriptscriptstyle ab}
  B_{\za} \,  H_{u}^a \tilde{L}_\za^b +
\epsilon_{\!\scriptscriptstyle ab} 
\left[ \, A^{\!\scriptscriptstyle U}_{ij} \,
\tilde{Q}^a_i H_{u}^b \tilde{U}^{\scriptscriptstyle C}_j
+ A^{\!\scriptscriptstyle D}_{ij}
H_{d}^a \tilde{Q}^b_i \tilde{D}^{\scriptscriptstyle C}_j
+ A^{\!\scriptscriptstyle E}_{ij}
H_{d}^a \tilde{L}^b_i \tilde{E}^{\scriptscriptstyle C}_j   \,
\right] + {\rm h.c.}\nonumber \\
&+&
\epsilon_{\!\scriptscriptstyle ab}
\left[ \,  A^{\!\scriptscriptstyle \lambda^\prime}_{ijk}
\tilde{L}_i^a \tilde{Q}^b_j \tilde{D}^{\scriptscriptstyle C}_k
+ \frac{1}{2}\, A^{\!\scriptscriptstyle \lambda}_{ijk}
\tilde{L}_i^a \tilde{L}^b_j \tilde{E}^{\scriptscriptstyle C}_k
\right]
+ \frac{1}{2}\, A^{\!\scriptscriptstyle \lambda^{\prime\prime}}_{ijk}
\tilde{U}^{\scriptscriptstyle C}_i  \tilde{D}^{\scriptscriptstyle C}_j
\tilde{D}^{\scriptscriptstyle C}_k  + {\rm h.c.}
\nonumber \\
&+&
 \tilde{Q}^\dagger \tilde{m}_{\!\scriptscriptstyle {Q}}^2 \,\tilde{Q}
+\tilde{U}^{\dagger}
\tilde{m}_{\!\scriptscriptstyle {U}}^2 \, \tilde{U}
+\tilde{D}^{\dagger} \tilde{m}_{\!\scriptscriptstyle {D}}^2
\, \tilde{D}
+ \tilde{L}^\dagger \tilde{m}_{\!\scriptscriptstyle {L}}^2  \tilde{L}
  +\tilde{E}^{\dagger} \tilde{m}_{\!\scriptscriptstyle {E}}^2
\, \tilde{E}
+ \tilde{m}_{\!\scriptscriptstyle H_{\!\scriptscriptstyle u}}^2 \,
|H_{u}|^2
\nonumber \\
&& + \frac{M_{\!\scriptscriptstyle 1}}{2} \tilde{B}\tilde{B}
   + \frac{M_{\!\scriptscriptstyle 2}}{2} \tilde{W}\tilde{W}
   + \frac{M_{\!\scriptscriptstyle 3}}{2} \tilde{g}\tilde{g}
+ {\rm h.c.}\; ,
\label{soft}
\end{eqnarray}
where we have separated the R-parity conserving ones from the
RPV ones ($H_{d} \equiv \hat{L}_0$) for the $A$-terms. Note that
$\tilde{L}^\dagger \tilde{m}_{\!\scriptscriptstyle \tilde{L}}^2  \tilde{L}$,
unlike the other soft mass terms, is given by a
$4\times 4$ matrix. Explicitly,
$\tilde{m}_{\!\scriptscriptstyle {L}_{00}}^2$ is
$\tilde{m}_{\!\scriptscriptstyle H_{\!\scriptscriptstyle d}}^2$
of the MSSM case while
$\tilde{m}_{\!\scriptscriptstyle {L}_{0k}}^2$'s give RPV mass mixings. 

Details of the tree-level mass matrices for all fermions and scalars are summarized in
Ref.\cite{GSSM}. For the analytical appreciation of many of the results, approximate
expressions of all the RPV mass mixings are very useful. The expressions are available
from perturbative diagonalization of the mass matrices\cite{GSSM}.

\section{The Quark Loop Contributions}
We perform calculations of the one-loop EDM diagrams using mass
eigenstates with their effective couplings. The approach frees our
numerical results from the mass-insertion approximation more commonly
adopted in the type of calculations, while analytical discussions
based of the perturbative diagonalization formulae help to trace the
major role of the RPV couplings, especially those of the bilinear
type.  The interesting class of one-loop contributions largely overlooked
by other authors on RPV physics consist of a bilinear-trilinear
parameter combinations. The bilinear parameters come into play through mass mixings
induced among the fermions and scalars (slepton and Higgs states), while
the trilinear parameter enters a effective coupling vertex.  The basic
features are the same as those reported in the studies of the various
related processes\cite{nedm,bsg,nedm-mssm}.  Among the latter, our recently
available reports on \bsg\ \cite{bsg} are particularly noteworthy, for comparison. 
The $d$ quark dipole plays a more important role over that of the $u$ quark, when RPV 
contributions are involved. The \bsg\ diagram is, of course, nothing other than a flavor
off-diagonal version of a down-sector quark dipole moment diagram.
 
The SM Yukawa couplings are real and flavor diagonal, not to say very
small for the $u$ and $d$ quarks.  However, the RPV analogs are mostly
not so strongly constrained in magnitudes\cite{bounds} and are sources
of flavor mixings. A trilinear $\lambda_{\za jk}^{\!\prime}$ coupling
couples a quark to another one, of the same or different charge, and a
generic scalar, neutral or charged accordingly.  The coupling together
with a SM Yukawa coupling at another vertex contributes to the EDMs
through some scalar mass eigenstates with RPV mass mixings involved,
as illustrated in Fig. \ref{dedm}. As the SM Yukawas are flavor diagonal,
the charged scalar loops contribute by invoking CKM mixings\footnote{
For the corresponding flavor off-diagonal transition moment, like
\bsg\ , scalar loop contributions do exist\cite{bsg}.}.  Note that
under the SVP adopted, the $\lambda_{\za jk}^{\!\prime}$ parameters
have quark flavor indices actually defined in the $d$-sector quark
mass eigenstate basis. Analytically, we have the formula
\bea \small
\label{eq:edm-formula}
\left( {\frac{d_{\scriptscriptstyle f}}{e}}\right) _{\!\!\phi ^{\mbox{-}}}=-{
\frac{\alpha _{\!\mbox{\tiny em}}}{4\pi \,\sin \!^{2}\theta _{\!
\scriptscriptstyle W}}}\;\sum_{m}^{\prime }\sum_{n=1}^{3}\,\mbox{Im}(
\wtl{\cal C}_{\ssc nmi}^{\ssc \!L}\wtl{\cal C}^{\ssc \!R \ast}_{\ssc nmi})
\frac{{M}_{\!\scriptscriptstyle f_{n}^{\prime }}}{M_{\!\scriptscriptstyle
\tilde{\ell}_{m}}^{2}}\;
\left[ ({\cal Q}_{\!f}-{\cal Q}_{\!f^{\prime
}})\;F_{4}\!\left( {\frac{{M}_{\!\scriptscriptstyle f_{n}^{\prime }}^{2}}{
M_{\!\scriptscriptstyle\tilde{\ell}_{m}}^{2}}}\right) -{\cal Q}_{\!f^{\prime
}}\;F_{3}\!\left( {\frac{{M}_{\!\scriptscriptstyle f_{n}^{\prime }}^{2}}{
M_{\!\scriptscriptstyle\tilde{\ell}_{m}}^{2}}}\right) \right] \;,
\end{eqnarray}
where, for $f=u$  ($i=1$), the coefficients 
$\wtl{\cal C}_{\ssc nmi}^{\! \ssc L,\ssc R}$ are defined by the interaction Lagrangian,
\be
{\cal L}^{u} = {g_{\scriptscriptstyle 2}} 
\overline{\Psi}(d_n)\Phi^{\dagger}(\phi_m^{{\mbox{-}}}) 
\left[ 
\widetilde{\cal C}^{\!\scriptscriptstyle L}_{\scriptscriptstyle nmi}
{1 - \gamma_{\scriptscriptstyle 5} \over 2}  + 
\widetilde{\cal C}^{\!\scriptscriptstyle R}_{\scriptscriptstyle nmi}
{1 + \gamma_{\scriptscriptstyle 5} \over 2}  \right]  
{\Psi}({u_{i}}) 
\ + \mbox{h.c.}\;,
\ee
\begin{eqnarray}
\wtl{\cal C}_{\ssc nmi}^{\ssc L \ast} &=&\frac{y_{d_{n}}}{g_{%
\scriptscriptstyle 2}}\,V_{\!\mbox{\tiny CKM}}^{in}{\cal D}_{2m}^{l^{\ast }}+\frac{\lambda
_{jkn}^{\!\prime \ast }}{g_{\scriptscriptstyle2}}\,V_{\!\mbox{\tiny CKM}}^{ik}{\cal D}%
_{(j+2)m}^{l^{\ast }}\;,  \nonumber \\
\wtl{\cal C }_{\ssc nmi}^{\ssc R \ast} &=&\frac{y_{u_{i}}}{g_{%
\scriptscriptstyle 2}}V_{\!\mbox{\tiny CKM}}^{in}\,{\cal D}_{1m}^{l^{\ast }}\;,
\end{eqnarray}
and, for $f=d$  ($i=1$), the coefficients 
$\wtl{\cal C}_{\ssc nmi}^{\! \ssc L,\ssc R}$ are defined by the interaction Lagrangian,
\be
{\cal L}^{d} = {g_{\scriptscriptstyle 2}} 
\overline{\Psi}(u_n)\Phi (\phi_m^{{\mbox{-}}}) 
\left[ 
\widetilde{\cal C}^{\!\scriptscriptstyle L}_{\scriptscriptstyle nmi}
{1 - \gamma_{\scriptscriptstyle 5} \over 2}  + 
\widetilde{\cal C}^{\!\scriptscriptstyle R}_{\scriptscriptstyle nmi}
{1 + \gamma_{\scriptscriptstyle 5} \over 2}  \right]  
{\Psi}({d_{i}}) 
\ + \mbox{h.c.}\;,
\ee
\begin{eqnarray}
\wtl{\cal C}_{\ssc nmi}^{\ssc L \ast}
&=&\frac{y_{u_{n}}}{g_{%
\scriptscriptstyle 2}}V_{\!\mbox{\tiny CKM}}^{ni^{\ast }}\,{\cal D}_{1m}^{l}\;,  
\nonumber \\
\wtl{\cal C }_{\ssc nmi}^{\ssc R \ast}
&=&\frac{y_{d_{i}}}{g_{%
\scriptscriptstyle 2}}\,V_{\!\mbox{\tiny CKM}}^{ni^{\ast }} {\cal D}^l_{2m}+
\frac{\lambda _{jki}^{\!\prime
}}{g_{\scriptscriptstyle 2}}\,V_{\!\mbox{\tiny CKM}}^{nk^{\ast }}{\cal D}
_{(j+2)m}^{l}\;,
\end{eqnarray}
with the $\sum_{m}^{\prime }$ denoting a sum over (seven) nonzero mass
eigenstates of the charged scalar; i.e., the unphysical Goldstone mode
is dropped from the sum, $D^{l}$ being the diagonalization matrix,
i.e., $D^{l\dag }M_{\!\scriptscriptstyle
E}^{2}\,D^{l}=\mbox{diag}\{\,M_{\!\scriptscriptstyle\tilde{\ell}
_{m}}^{2},m=1\,\mbox{--}\,8\,\}$ and $M_{\!\scriptscriptstyle
E}^{2}$ being $8\times 8 $ charged-slepton and Higgs mass-squared matrix .
Here, $i=1$ is the flavor index for the external quark while $n$ is
for the quark running inside the loop. Note that the unphysical
Goldstone mode is dropped from the scalar sum because it is rather a
part of the gauge loop contribution, which obviously is real and does
not affect the EDM calculation. 

In general, there also exist the contributions from neutral scalar
loop (involving the mixing of neutral Higgs with sneutrino in the
loop). These contributions lack the top Yukawa and top mass
effects that enhance the contributions of charged-slepton loop, and
hence less important. The formula is given as:
\be
\label{eq:neutral-scalar}
\left( {\frac{d_{\scriptscriptstyle
f}}{e}}\right)_{\!\!\phi^{\ssc 0}}=
-
\frac{\alpha _{\!\mbox{\tiny em}} {\cal Q}_f}{4\pi \,\sin \!^{2}\theta _{\!
\scriptscriptstyle W}}\;\sum_{m}^{\prime }\sum_{n=1}^{3}\,\mbox{Im}(
\wtl{\cal N}_{\ssc nmi}^{\ssc \!L}\wtl{\cal N}^{\ssc \!R \ast}_{\ssc nmi})
\frac{{M}_{\!\scriptscriptstyle f_{n}}}{M_{\!\scriptscriptstyle
\tilde{s}_{m}}^{2}}\;
\;F_{3}\!\left( {\frac{{M}_{\!\scriptscriptstyle f_{n}}^{2}}{
M_{\!\scriptscriptstyle\tilde{\ell}_{m}}^{2}}}\right) \;,
\ee
where, for $f=u$  ($i=1$), the coefficients $\wtl{\cal N}_{\ssc nmi}^{\! \ssc L,\ssc R}$ 
are defined by the interaction Lagrangian,
\be
{\cal L}^{u} = {g_{\scriptscriptstyle 2}} 
\overline{\Psi}(u_n)\Phi^{\dagger}(\phi_m^{{\ssc 0}}) 
\left[ 
\widetilde{\cal N}^{\!\scriptscriptstyle L}_{\scriptscriptstyle nmi}
{1 - \gamma_{\scriptscriptstyle 5} \over 2}  + 
\widetilde{\cal N}^{\!\scriptscriptstyle R}_{\scriptscriptstyle nmi}
{1 + \gamma_{\scriptscriptstyle 5} \over 2}  \right]  
{\Psi}({u_{i}}) 
\ + \mbox{h.c.}\;,
\ee
\bea
\widetilde{\cal N}^{\!\scriptscriptstyle L \ast}_{\scriptscriptstyle
nmi} & = & -\f{y_{u_i}}{g_2}\d_{in}\f{1}{\sqrt 2} \left[ {\cal
D}^s_{1m} - i {\cal D}^s_{6m} \right] \; ,\nonumber\\
\widetilde{\cal N}^{\!\scriptscriptstyle R \ast}_{\scriptscriptstyle
nmi} & = & -\f{y_{u_i}}{g_2}\d_{in}\f{1}{\sqrt 2} \left[ {\cal
D}^s_{1m} + i {\cal D}^s_{6m} \right]\;;
\eea
and, for $f=d$  ($i=1$), the coefficients 
$\wtl{\cal N}_{\ssc nmi}^{\! \ssc L,\ssc R}$ are defined by the
interaction Lagrangian,
\be
{\cal L}^{d} = {g_{\scriptscriptstyle 2}} 
\overline{\Psi}(d_n)\Phi (\phi_m^{\ssc 0}) 
\left[ 
\widetilde{\cal N}^{\!\scriptscriptstyle L}_{\scriptscriptstyle nmi}
{1 - \gamma_{\scriptscriptstyle 5} \over 2}  + 
\widetilde{\cal N}^{\!\scriptscriptstyle R}_{\scriptscriptstyle nmi}
{1 + \gamma_{\scriptscriptstyle 5} \over 2}  \right]  
{\Psi}({d_{i}}) 
\ + \mbox{h.c.}\;,
\ee
\bea
\widetilde{\cal N}^{\!\scriptscriptstyle L \ast}_{\scriptscriptstyle
nmi} & = & -\f{y_{u_i}}{g_2}\d_{in}\f{1}{\sqrt 2} \left[ {\cal
D}^s_{2m} - i {\cal D}^s_{7m} \right] 
-\f{\l^{\prime \ast}_{jin}}{g_2}\f{1}{\sqrt 2} \left[
{\cal D}^s_{(j+2)m} - i {\cal D}^s_{(j+7)m} \right] \nonumber\\
\widetilde{\cal N}^{\!\scriptscriptstyle R \ast}_{\scriptscriptstyle
nmi} & = & -\f{y_{d_i}}{g_2}\d_{in}\f{1}{\sqrt 2} \left[ {\cal
D}^s_{2m} + i {\cal D}^s_{7m} \right]
-\f{\l^{\prime}_{jni}}{g_2}\f{1}{\sqrt 2} \left[
{\cal D}^s_{(j+2)m} + i {\cal D}^s_{(j+7)m} \right]
\; .
\eea
with the $\sum_{m}^{\prime }$ denoting a sum over (10) nonzero mass
eigenstates of the neutral scalar; i.e., the unphysical Goldstone mode
is dropped from the sum, $D^{s}$ being the diagonalization matrix for
the $10\times 10$ mass matrix for neutral scalars (real and symmetric,
written in terms of scalar and pseudo-scalar parts). Again $i=1$ is the
flavor index for the external quark while $n$ is for the quark running
inside the loop.

\section{Results}
Quark EDMs involve violation of CP but not $R$-parity. Thus R-parity
violating parameters should come in combinations that conserve
R-parity. From an inspection of formulae it is clear that the
contributions from two $\l^{\prime}$ couplings cannot lead to EDM as
these violate lepton number by two units which has to be compensated
by Majorana like mass insertions for neutrino or sneutrino
propagators. If one is willing to admit more than two $\l^{\prime}$ to
be non-zero, than in principle one can have contributions to EDM from the
Majorana like mass insertions but these would be highly suppressed. With
only two RPV couplings, the only other possibility is to have a
combination of trilinear and a bilinear ($\m_i$, $B_i$ or 
${\tilde m}^2_{{\ssc L}_{\ssc 0i}}$) RPV couplings in such a way that lepton number is
conserved. However not all the three bilinears mentioned above are
independent as they are related by tadpole relation in the
single VEV parametrization \cite{GSSM}:
\be
B_i \tan\! \beta = {\tilde m}^2_{{\ssc L}_{\ssc 0i}} + \m^*_{\ssc 0}\m_i ~.
\ee
Using the above tadpole equation we eliminate 
${\tilde m}^2_{{\ssc L}_{\ssc 0i}}$ in favor of $\m_i$ and $B_i$. Contributions from
the combination $\m^{\ast}_i\l^{\prime}_{ijk}$, through squark loops, had been 
extensively studied in \cite{nedm} in detail. Here, we shall focus on the $B^{\ast}_i 
\l^{\prime}_{ijk}$ kind of combination as illustrated in Fig.\ref{dedm}.
Such a combination can contribute through charged-scalar
(charged-slepton, charged Higgs mixing) loop or neutral scalar
(sneutrino, neutral Higgs mixing) loop. The fermions running inside the loops are
quarks, instead of the gluon or a colorless fermion as in the case of the squark loops.
Before we discuss the numerical results it would be worth-while to discuss the analytical 
results which can then be compared with numerical results.

Let us focus on RPV part of 
$ \mbox{Im}(\wtl{\cal C}_{\ssc nmi}^{\ssc \!L}\wtl{\cal C}^{\ssc \!R \ast}_{\ssc nmi})$,
in Eq.(\ref{eq:edm-formula})  for the case of $d$ quark EDM. It is given as:
\be
\label{cs-d}
\mbox{Im}(
\wtl{\cal C}_{\ssc nm1}^{\ssc \!L}\wtl{\cal C}^{\ssc \!R \ast}_{\ssc
nm1})_{\ssc RPV}  = \mbox{Im}\left[
(y_{u_n}\vckm^{n1}{\cal D}^{l\ast}_{1m})~\times~(\l^{\prime}_{jk1}
\vckm^{nk\ast}{\cal D}^{l}_{(j+2)m})\right].
\ee
For the $u$ quark dipole, we have
\be
\label{cs-u}
\mbox{Im}(
\wtl{\cal C}_{\ssc nm1}^{\ssc \!L}\wtl{\cal C}^{\ssc \!R \ast}_{\ssc
nm1})_{\ssc RPV} = \mbox{Im}\left[
(\l^{\prime}_{jkn}\vckm^{1k\ast} {\cal D}^l_{(j+2)m})
~\times~(y_{u}\vckm^{1n}
{\cal D}^{l\ast}_{1m})
\right].
\ee
Interestingly, the entries of the slepton-Higgs diagonalizing matrix involved are
the same in both terms above.  Bilinear RPV terms are
hidden inside the slepton-Higgs diagonalization matrix elements. To
see the explicit dependence on bilinear RPV terms let us look at the
diagonalizing matrix elements of charged-slepton Higgs mass matrix,
${\cal D}^l_{(j+2)m}{\cal D}^{l\ast}_{1m}$, more closely.  When summed over
the index $m$, it of course gives zero, owing to unitarity. This is possible only for
the case of exact mass degeneracy among the scalars when loop functions factor out
in the sum over $m$ scalars. In fact, the final result involves a double summation
over the $m$ scalar and $n$ fermion (quark) mass eigenstates. Either the unitarity 
constraint over the former sum or the GIM cancellation over the latter predicts a null
result whenever the mass dependent loop functions 
$F_{4}\!\left( {\frac{{M}_{\!\scriptscriptstyle f_{n}^{\prime }}^{2}}{
M_{\!\scriptscriptstyle\tilde{\ell}_{m}}^{2}}}\right)$  and 
$F_{3}\!\left( {\frac{{M}_{\!\scriptscriptstyle f_{n}^{\prime }}^{2}}{
M_{\!\scriptscriptstyle\tilde{\ell}_{m}}^{2}}}\right)$  can be factored out of
the corresponding summation due to mass degeneracy. In reality
however, these elements are multiplied by non-universal loop functions giving
non-zero EDM. Restricting to first order in perturbation expansion of
the mass-eigenstates, ${\cal D}^l_{(j+2)m}{\cal D}^{l\ast}_{1m}$ is
non-zero for $m=2$ and $m=j+2$, both giving similar dependence on RPV
but with opposite sign
($m=1$ is the unphysical Goldstone state that is dropped from the sum here). 
For $m=2$ one obtains \cite{GSSM}:
\be
\label{diag1}
{\cal D}^l_{(j+2)2}\;{\cal D}^{l\ast}_{12} \sim \f{ B^{\ast}_j}{M^2_s} 
 \times \mathrm{O}(1)
\ee
Here, $M^2_s$ denotes the difference in the relevant diagonal entries (generic
mass-squared parameter of the order of soft mass scale) in
charged-slepton Higgs mass matrix. 
To first order in perturbation
expansion there is no contribution to EDM from $\m_i$. If one goes to
second order in perturbation expansion, one gets a contribution from
the term $m = j+5$ which is given as \cite{GSSM}:
\be
\label{mui}
{\cal D}^l_{(j+2)(j+5)}\;{\cal D}^{l\ast}_{1(j+5)} \sim \f{\m_j^{\ast} m_j}{M_s^2}
\times \left[\f{(A^{\ast}_e - \m_{\ssc 0} \tan \b)m_j}{M^2_s} - 
\f{\sqrt{2}M_{\ssc W} \sin \b (\m_k \l^{\ast}_{kjj})}{g_2 M^2_s}\right]
\ee
There are a few important things to be noted here:
\begin{enumerate}
\item 
Notice that $\m_i$ enter at second order in perturbation
expansion. Moreover they are accompanied by
corresponding charged-lepton mass-squared and hence $\m^{\ast}_i \l^{\prime}_{ijk}$
contributions are always suppressed (later below we will elaborate on
this). Also notice that in principle trilinear parameter $\l_{kjj}$
also contribute through $\m_k \l^{\ast}_{kjj}$. 
But they have to be present in addition to the trilinear
parameter $\l^{\prime}_{ijk}$, thus making it a fourth-order effect in
perturbation and hence negligible. Thus, we will focus on $B^{\ast}_i
\l^{\prime}_{ijk}$ effects which are interesting. 

\item Even if all RPV parameters are real, CKM phase in conjunction 
with real RPV parameters could still induce EDM. As we will soon see,
this could be sizable.
\item 
It is clear that $d$ quark EDM receives much larger contribution
owing to top-Yukawa and proportionality to top mass. There are nine
 trilinear RPV couplings $\l^{\prime}_{ij1}$ that contribute to
$d$ quark EDM. $\l^{\prime}_{i31}$ has the largest impact owing to
the least CKM suppression.
\item 
All the 27 trilinear RPV couplings ($\l^{\prime}_{ijk}$) contribute to $u$ quark EDM. 
However, the absence of enhancement from the top mass in the loop
and the uniform proportionality to up-Yukawa considerably weakens the type of
contribution to $u$ quark EDM.
\item 
There is no RPV neutral scalar loop contribution to $u$ quark EDM.
However, there are one-loop contributions to $d$ quark EDM
from the neutral-Higgs sneutrino mixing due to the combination of
$B_i \l^{\prime}_{i11}$ with Majorana like mass-insertion in the
loop\footnote{The  Majorana like mass-insertion can be considered a result
of the non-zero $B_i$. It manifest itself in our exact mass eigenstate calculation
as a mismatch between the corresponding scalar and pseudo-scalar parts
complex ``sneutrino" state which would otherwise have contributions canceling
among themselves. With the non-zero $B_i$, the involved diagram is one with
two $\l^{\prime}_{i11}$ coupling vertices and a internal $d$ quark. Hence,
the type of contribution is possible only with the single  $\l^{\prime}$ coupling.
}. From the EDM formula, it is clear that this would be about the
same magnitude as the $u$ quark EDM due to charged-Higgs charged
slepton mixing and hence highly suppressed. 
\end{enumerate}

From the above analytical discussion, we illustrated clearly how the various
combinations of trilinear and bilinear RPV parameters contribute to neutron EDM. 
We now focus on the numerical results.  In order to focus on
individual contributions we keep a pair of RPV coupling (a bilinear
and a trilinear) to be non-zero at a time to study its impact. We have
chosen all sleptons and down-type Higgs to be 100 GeV (up-type Higgs
mass and $B_{0}$ being determined from electroweak symmetry
breaking conditions), $\m_{\ssc 0}$ parameter to be -300 GeV, 
$A$ parameter at the value of 300 GeV, and $\tan\!\b = 3$ (we will show 
the impact of some parameter variations below). Since the $\l^{\prime}$
couplings are on the same footing as the standard Yukawa couplings,
they are in general complex.  We admit a phase of $\pi/4$ for
non-zero $\lambda^{\prime}$ couplings while still keeping the CKM
phase. The phases of the other R-parity conserving parameters are switched
off. Since it is the complex phase of the $B^{\ast}_i \l^{\prime}_{ijk}$ product
that is importance, we put the phase for $B_i$ to be zero without loss
of generality.  We do not assume any hierarchy in the sleptonic
spectrum. Hence, it is immaterial which of the bilinear parameter $B_i$ is
chosen to be non-zero. We choose $B_3$ to be non-zero but all the
results hold good for $B_1$ and $B_2$ as well.

In Fig.(\ref{b3-331}) we have plotted the neutron EDM versus the most
important combination $|B_3^{\ast}\l^{\prime}_{331}|$ normalized by
$\m_{\ssc 0}^2$. As mentioned above, it can be clearly seen that the $d$ quark
EDM contribution from the charged-slepton loop (green circles) dominates
over the $u$ quark EDM (blue triangles). The experimental upper bound is
shown with the pink horizontal dotted line and the QCD corrected total
neutron EDM is shown in red squares. From the plot one can extract an
upper bound of $4.0\cdot 10^{-5}$ on the combination of couplings
$|B_3^{\ast}\l^{\prime}_{331}|/\m_{\ssc 0}^2$. However this bound corresponds to a
fixed value of relative phase ($\pi/4$) in the RPV parameter
combination) as well as other supersymmetric parameters. Later in the
table \ref{bounds}, we show the effects of variation in the major
parameters. With a similar value of input parameters we obtain an upper  
bound of $2.2\cdot 10^{-4}$ for $|B_3^{\ast}\l^{\prime}_{321}|/\m_{\ssc 0}^2$ and
$8.4\cdot 10^{-4}$ for $|B_3^{\ast}\l^{\prime}_{311}|/\m_{\ssc 0}^2$. The hierarchy in
these bounds is due to the relevant CKM factors. 

Fig.(\ref{contour}) shows the contours of neutron EDM in the plane of
magnitudes for the couplings $\l^{\prime}_{331}$ and $B_3$ with the relative phase
fixed at $\pi/4$. The contour for present experimental bound is shown
in pink dotted line. The plot shows that  a sizable region of the parameter
space are ruled out. Successive contours show smaller values of
neutron EDM with the smallest being $10^{-27}$ $e$ cm.

So far we have kept certain parameters like the phase of the RPV
combination, the $\mu_{\ssc 0}$ parameter and the slepton mass spectrum fixed.  To
get a better understanding of the allowed regions in the overall parameter space, 
let us focus  on variations of the parameters one at a time. 
In Fig.(\ref{m3}) we have plotted neutron EDM versus the slepton
mass parameter ${\tilde m}_{L} = {\tilde m}_{E}$ (with $|B_3| = 200
~\mbox{GeV}^2$, $|\l^{\prime}_{331}| = .05$ and relative phase of $\pi/4$). 
As expected the neutron EDM falls with the increasing slepton mass.
More careful checking reveals that the result is sensitive only to 
one slepton mass parameter,
the mass of the third $L$-handed slepton here. It is also easy to understand from our analytical 
formulas that the dominating contributions  among the various scalar mass eigenstates for the 
case of  $B_i^{\ast}\l^{\prime}_{ijk}$ come from the $i$-th $L$-handed slepton and the Higgs.
In Fig.(\ref{mu}) we have shown the variation of neutron EDM with the
$\m_{\ssc 0}$ parameter. Although parameter $\m_{\ssc 0}$ does not directly figure in
the EDM formula, its influence is felt through the Higgs
spectrum. Larger $\m_{\ssc 0}$ leads to heavier Higgs spectrum which suppresses
the EDM contribution.  As our analytical formulas show there is no
strong dependence on $\tan\! \b$ (we have checked this numerically as
well) and hence we have kept $\tan\! \b =3$ fixed in all the plots.

So far we have discussed only the combination
$ B_i^{\ast}\l^{\prime}_{i31}$ for the purpose of illustration. In table
\ref{bounds}, we list the bounds on the nine combinations 
$|B_i^{\ast}  \l^{\prime}_{ij1}|$ normalized by $(100~\mbox{GeV})^2$. It is
interesting to note the difference in bounds for case (a) and (b) for
coupling combination $|B_i^{\ast} \l^{\prime}_{i31}|$ and
$|B_i^{\ast}\l^{\prime}_{i21}|$. The inputs for Case (b) are identical to
case (a) except for RPV phase being zero for case (b). Case (b) thus
relies solely on CKM phase. Interestingly the bound for
$|B_i^{\ast}\l^{\prime}_{i31}|$ changes very marginally from case (a) to
(b) whereas the bound for $|B_i^{\ast} \l^{\prime}_{i21}|$ weakens by
about an order of magnitude. To understand this difference in
behavior of $|B_i^{\ast}\l^{\prime}_{i31}|$ and $|B_i^{\ast}\l^{\prime}_{i21}|$, 
with and without a complex phase in the RPV couplings, we have plotted in
Fig. \ref{scatter} the allowed region in the plane of 
relative phase of $B^{\ast}_3 \l^{\prime}_{331}$ and the
$|B^{\ast}_3\l^{\prime}_{331}|$ (left) and 
relative phase of $B^{\ast}_3 \l^{\prime}_{321}$ and the
$|B^{\ast}_3 \l^{\prime}_{321}|$ (right). One can see that
the bound for $|B^{\ast}_3 \l^{\prime}_{331}|$ is about the same for
relative phase in $B^{\ast}_3 \l^{\prime}_{331}$ of 0 or $\pi/4$ but
the bound for $|B^{\ast}_3 \l^{\prime}_{321}|$ strengthens by
about an order of magnitude as the phase in the RPV coupling increases from 0 to 
$\pi/4$ suggesting a collaborative effect between the CKM phase and the RPV phase. 
For the case of  $B_i^{\ast} \l^{\prime}_{i11} $ there is no CKM phase involved.
In the table we have fixed the sign of $\m_{\ssc 0}$ parameter to be negative. In
the Fig.~\ref{mu} it is seen that absolute value of neutron EDM
is symmetric with respect to sign of $\m_{\ssc 0}$ parameter and hence positive
$\m_{\ssc 0}$ should give identical bounds.
The variation of bounds in table~\ref{bounds} with changes in parameters
$\m_{\ssc 0}$ and the slepton and Higgs mass very much follows the
pattern found in plots discussed earlier.
\TABULAR{llllllll}{\hline\hline
\multicolumn{5}{c}{{\bf parameter values}} & \multicolumn{3}{c}{{\bf Normalized bounds}}\\
&$\m_{\ssc 0}$ & ${\tilde m}_{\ssc L}$ & $m_{{\ssc H}_d}$ &
RPV  & $\f{|B_i^{\ast}\cdot \l^{\prime}_{i31}|}{(100~\mbox{GeV})^2} $ &
$\f{|B_i^{\ast}\cdot \l^{\prime}_{i21}|}{(100~\mbox{GeV})^2}$ & $\f{|B_i^{\ast}\cdot
\l^{\prime}_{i11}|}{(100~\mbox{GeV})^2}$\\ 
&~&~&~&phase&~&~\\\hline
(a)&-100 GeV & 100 GeV & 100 GeV
& $\pi/4$ & $1.6\cdot 10^{-4}$ & $4.8\cdot 10^{-4}$ & $2.0\cdot
10^{-3}$\\ 
(b)&-100 GeV & 100 GeV & 100 GeV & 0 & $1.8\cdot 10^{-4}$ &
$4.5\cdot 10^{-3}$ & N.A. \\ 
(c)&-400 GeV & 100 GeV & 100 GeV & $\pi/4$ &
$5.0\cdot 10^{-4}$ & $3.2\cdot 10^{-3}$ & $1.1\cdot 10^{-2}$\\ 
(d)&-800 GeV & 100 GeV & 100 GeV & $\pi/4$ & $1.7\cdot 10^{-3}$ & $1.0\cdot
10^{-2}$ & $3.7\cdot 10^{-2}$\\ 
(e)&-100 GeV & 400 GeV & 100 GeV & $\pi/4$
& $6.5\cdot 10^{-4}$ & $5.0\cdot 10^{-3}$ & $1.8\cdot 10^{-2}$\\ 
(f) &-100
GeV & 800 GeV & 100 GeV & $\pi/4$ & $2.0\cdot 10^{-3}$ & $1.7\cdot
10^{-2}$ & $5.8\cdot 10^{-2}$\\ 
(g)&-100 GeV & 100 GeV & 300 GeV & $\pi/4$
& $4.9\cdot 10^{-4}$ & $2.1\cdot 10^{-3}$ & $7.5\cdot 10^{-3}$\\ 
(h)&-100
GeV & 100 GeV & 600 GeV & $\pi/4$ & $1.1\cdot 10^{-3}$ & $6.8 \cdot
10^{-3}$ & $2.4\cdot 10^{-2}$\\
\hline\hline }{Here we list the normalized upper bounds for several
combinations of bilinear and trilinear RPV parameters, with some
variation in the input parameters. The bounds essentially depend on the
values of parameters like ${\tilde m}_{\ssc L},\m_{\ssc 0}$ and
$m_{{\ssc H}_d}$ ($m_{{\ssc H}_u}$ and $B_{0}$ being determined from EW
symmetry breaking condition). $\tan\! \b$ has been
kept fixed at 3 as EDM has a very mild dependence on $\tan\! \b$. 
\label{bounds}}

Before we conclude, we would like to briefly comment on two things. In the
table~\ref{bounds}, we have only mentioned the bounds for the
combinations $| B_i^{\ast} \l^{\prime}_{ij1}|$, whereas earlier
in the text we did mention that in principle all the twenty-seven
$\l^{\prime}_{ijk}$ together with bilinear couplings do contribute to
the neutron EDM. The couplings other than $\l^{\prime}_{ij1}$ contribute
to $u$ quark EDM and all the contributions are proportional to
up-Yukawa (in contrast to the presence of a contribution proportional to top-Yukawa  
for the $d$ quark EDM). Strength of the corresponding contributions is substantially 
weaker (typically by 5 to 6 orders of magnitude) and hence do not lead to
meaningful bounds. The other thing is about the possible $\mu_i^{\ast}
\cdot \l^{\prime}_{ijk}$ contributions. It can be seen in
Eq.(\ref{mui}) that these are second order in perturbation and also
accompanied by lepton mass $\m_i$.
Thus, a typical $\m_i^{\ast} \l^{\prime }_{ijk}$
contribution is substantially weaker than the corresponding
$B_i^{\ast}  \l^{\prime }_{ijk}$ contribution. For instance,
with the similar inputs for other SUSY parameters as described above,
if one takes $\mu_{3} = 10^{-3}$ GeV (dictated by requirement of sub-ev
neutrino masses), $\l^{\prime}_{331} = .05$ and a relative phase of
$\pi /4$, one obtains neutron EDM of $6.0 \cdot 10^{-32}$ which is
about six orders of magnitude smaller than the present experimental
bound on neutron EDM. If one goes by the upper bound on the mass of
$\n_{\tau}$ of 18.2 MeV \cite{nu_mass}, $\mu_{\ssc 3}$ could be as large as 7
GeV for $\tan\! \b = 2$ and sparticle mass $\sim 300$ GeV 
\cite{ru2}. For $\m_{\ssc 3}= 1$
GeV we obtain neutron EDM of $6.1 \cdot 10^{-29}$, still about three
orders of magnitude smaller than the experimental bound. 
These numbers can be easily compared with the $\m^{\ast}_i
\l^{\prime}_{i11}$ contributions to the $d$ quark EDM through
chargino loop in Table~1 of ref.\cite{nedm}. There the corresponding
contribution (with $\m_{\ssc 3} = 1$ GeV) is much weaker (of order
$10^{-32}$) as it lacks the top Yukawa and top mass enhancements. In
the same table one finds that corresponding contribution to gluino
loop is much stronger (of order $10^{-25}$) due to proportionality to
gluino mass and strong coupling constant. In the light of above reasons
one can appreciate that the contributions due to soft parameter $B_i$
are far more dominating in the present scenario of quark loop
contributions.

\section{Conclusions}
We have made a systematic study of the influence of bilinear and
trilinear RPV couplings on the neutron EDM. A combination of bilinear
and trilinear RPV is the only way RPV parameters can contribute at one
loop level. RPV couplings contribute to both, $u$ quark as well as $d$
quark EDM. Such a contribution could come from charged-slepton Higgs
mixing loop (contributing to both $u$ quark as well as $d$ quark EDM)
or sneutrino-Higgs mixing loop (viable only for combination of nonzero
$B_i \l^{\prime}_{i11}$ giving Majorana-like scalar mass insertion)
contributing to only $d$ quark EDM. The class of diagrams all have a quark as
the fermion running inside the loop. In our analytical expressions
obtained based on perturbative diagonalization of the scalar mass-squared matrices, 
we demonstrated that charged-slepton Higgs mixing loop
contribution to $d$ quark EDM far dominates the other contributions
due to a diagram with the top-quark in the loop. In our numerical exercise we have
obtained robust bounds on the combinations $\f{|B_i^{\ast}\cdot
\l^{\prime}_{ij1}|}{(100~\mbox{GeV})^2} $ for $i,j=1,2,3$ that have not been
reported before. Even if the RPV couplings are real, they could still
contribute to neutron EDM via CKM phase. For some cases CKM phase induced
contribution is as strong as that due to an explicit complex phase in the RPV 
couplings. There also exist contributions involving $\m_i^{\ast}
\l^{\prime}_{ijk}$. However these are higher order effects which are
further suppressed by proportionality to charged lepton mass. Since
$\m_i$ are expected to be very small (of order $10^{-3}$ GeV) for
sub-eV neutrino masses, such contributions are highly suppressed.
Our results presented here make available a new set of interesting bounds
on combinations of RPV parameters.

\FIGURE{\epsfig{file=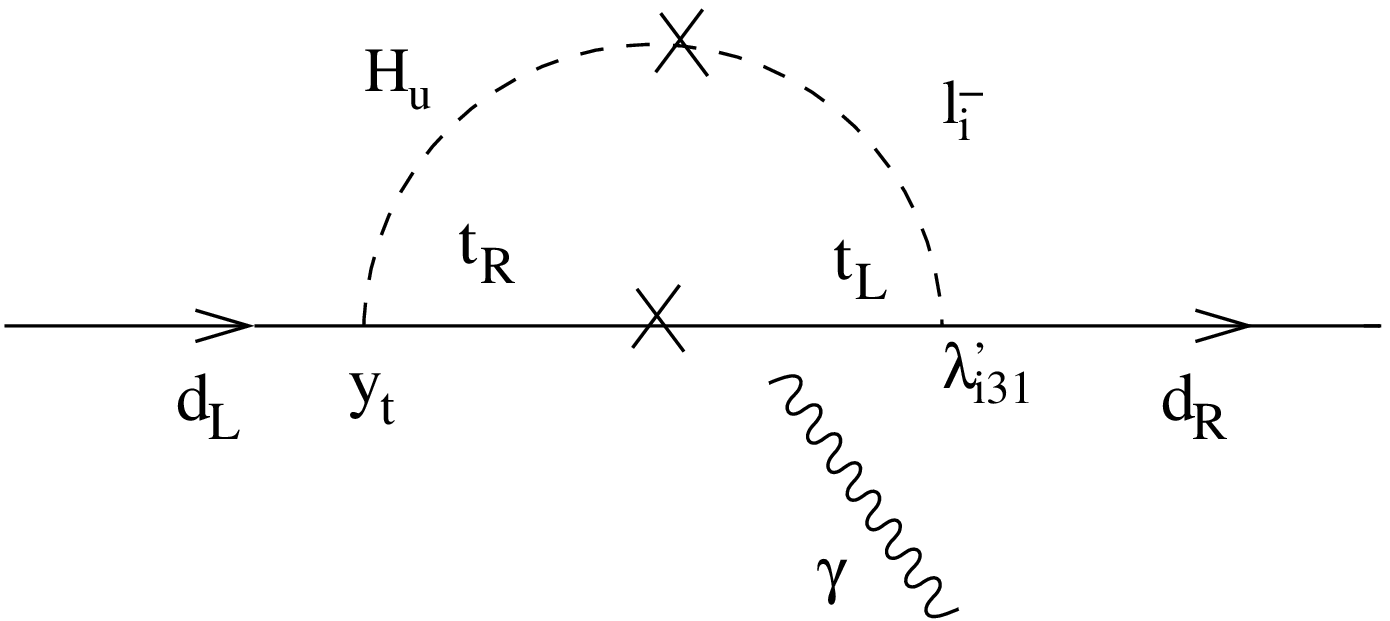,width=10cm}
\caption{\label{dedm}$d$ quark EDM due to $B^{\ast}_i
\l^{\prime}_{i31}$ combination. Due to top Yukawa and top-mass
dependence this is the most dominant contribution.}}
\FIGURE{\epsfig{file=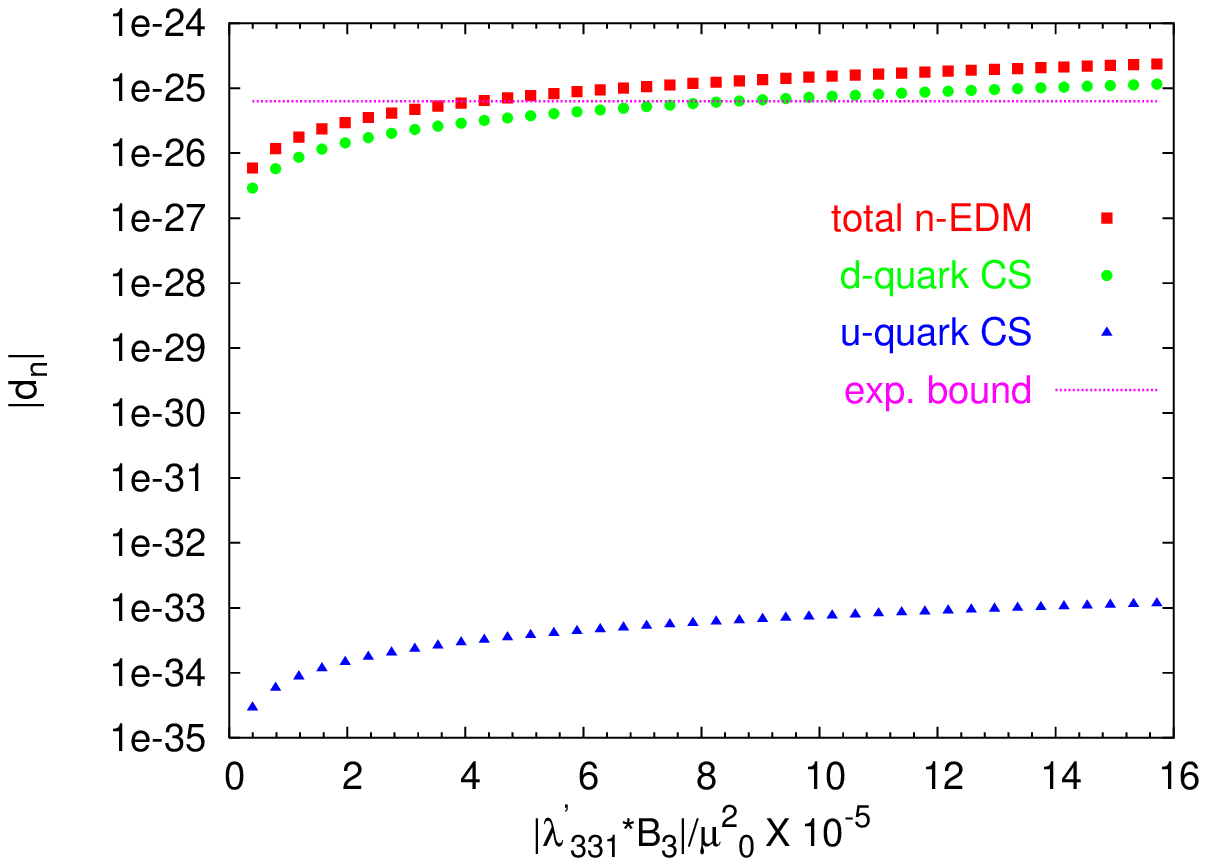,width=10cm}
\caption{\label{b3-331}(color online)
The neutron EDM (in units of e.cm) versus the combination $|B_3^{\ast}\l^{\prime}_{331}|$
normalized by $\m_{\ssc 3}^2$. The charged scalar contribution (green dotted
line) to the $d$ quark EDM is clearly seen to dominate over the
charged-scalar $u$ quark EDM (blue dotted line). The total QCD
corrected neutron EDM is shown in red line. The experimental
upper bound is shown in pink dotted line.}}
\FIGURE{\epsfig{file=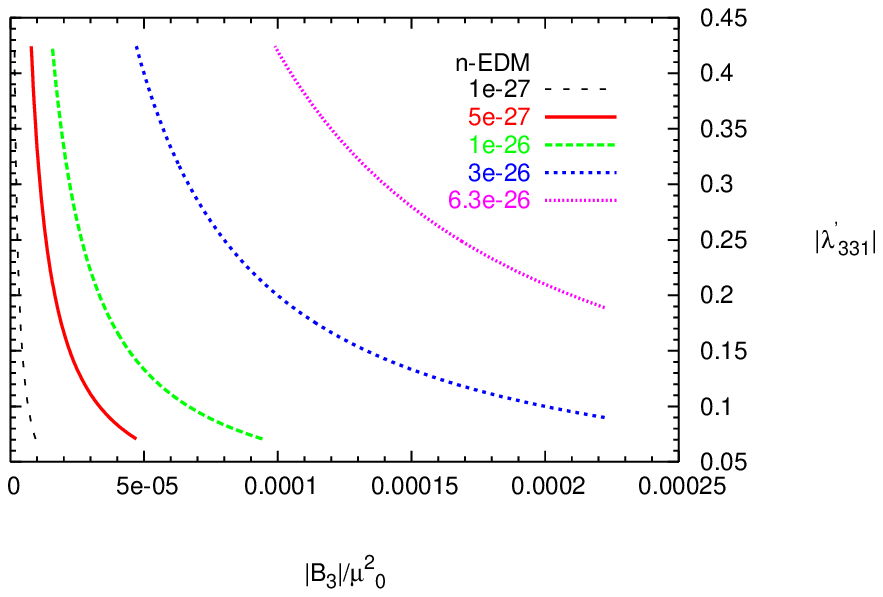}\caption{\label{contour} 
(color online) A contour plot for various values of neutron
EDM in e.cm in the plane of real $\l^{\prime}_{331}$ and $B_3$. The
red line is the present experimental bound for neutron EDM.}}
\FIGURE{\epsfig{file=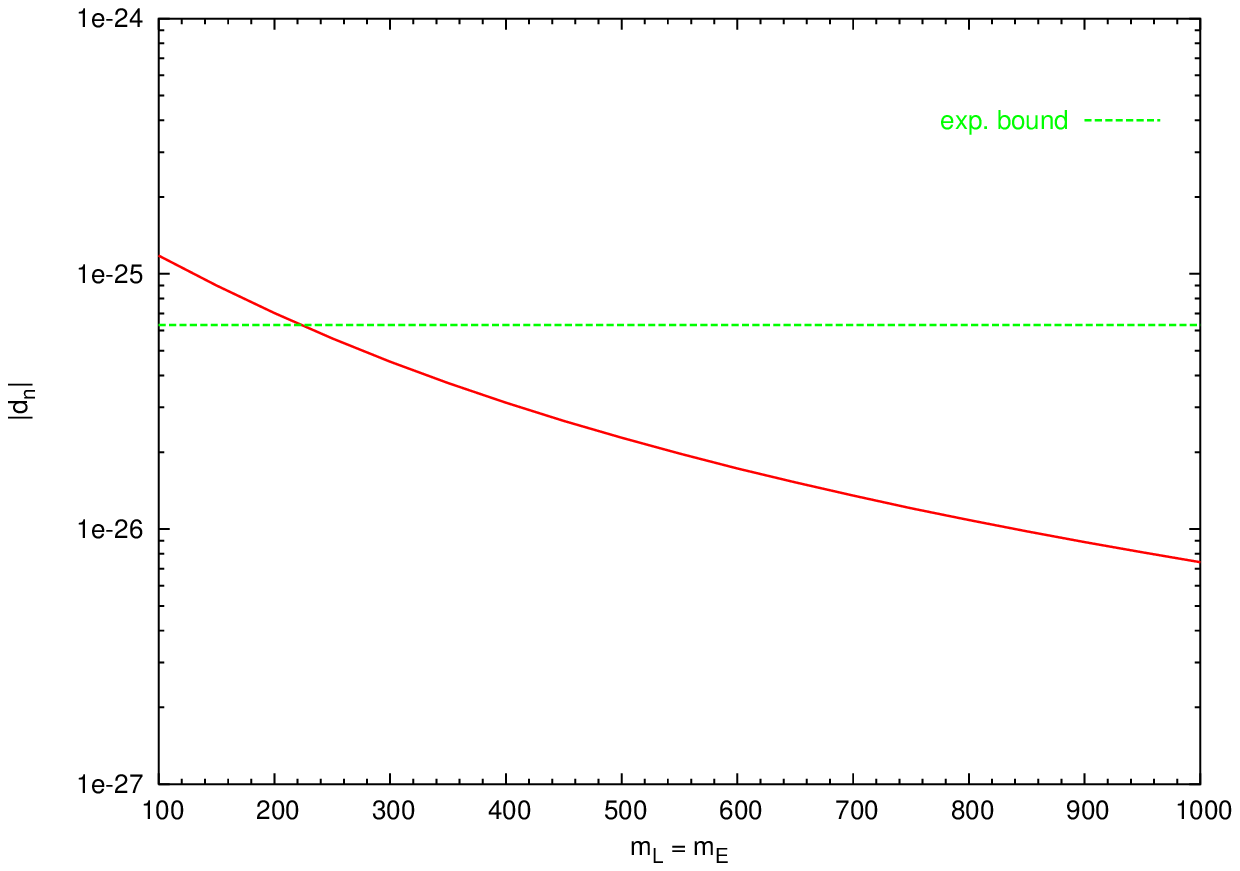}\caption{\label{m3} Neutron EDM vs. 
the slepton
mass parameter ${\tilde m}_{L} = {\tilde m}_{E}$ (with $|B_3| = 200
~\mbox{GeV}^2$, $|\l^{\prime}_{331}| = .05$ and relative phase of $\pi/4$).
Horizontal line is the experimental bound.}}
\FIGURE{\epsfig{file=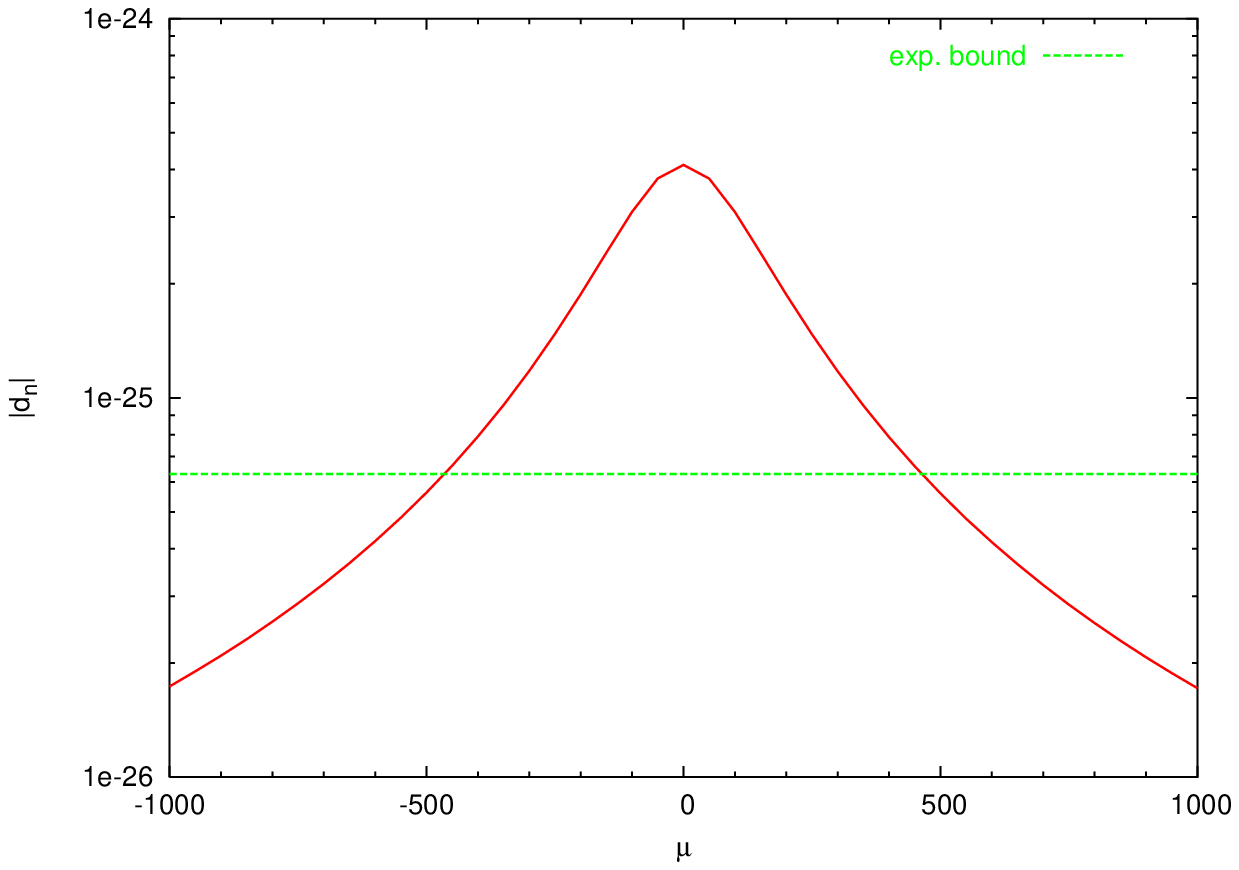}\caption{\label{mu} Neutron EDM vs. $\m_{\ssc 0}$ parameter with
$|B_3| = 200 ~\mbox{GeV}^2$, $|\l^{\prime}_{331}| = .05$ and relative
phase $\pi/4$. Horizontal line is the experimental bound.}}
\FIGURE{
\parbox{5.8in}{  
\hskip 0.03in
\epsfig{file=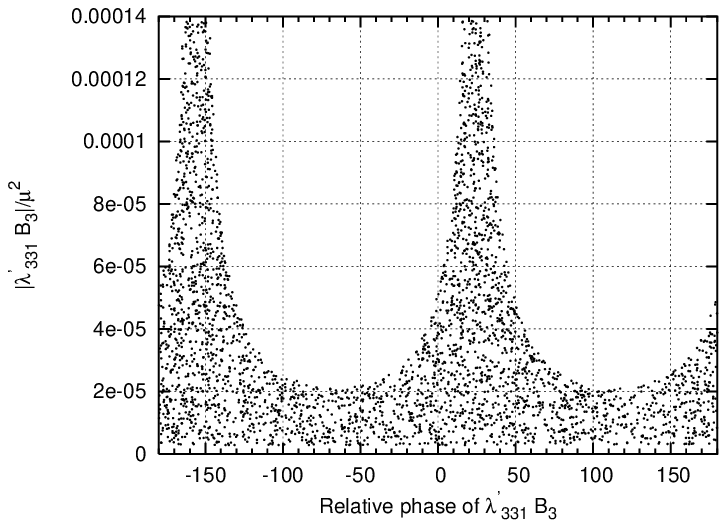,width=7cm}
\hskip 5mm   
\epsfig{file=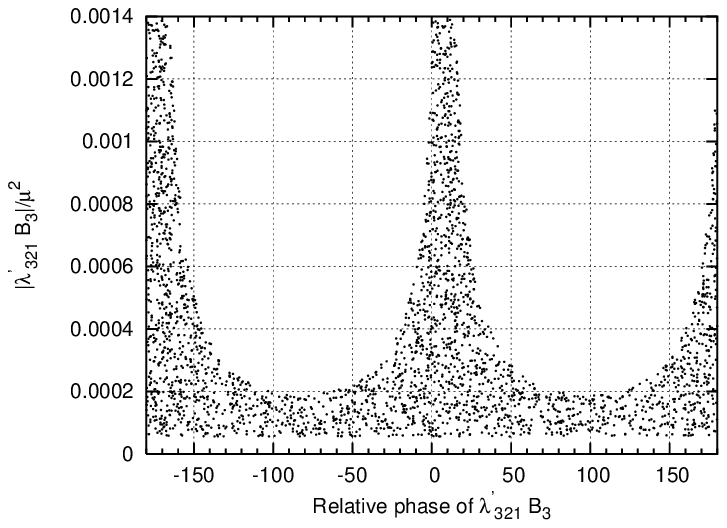,width=7cm}}
\caption{
Dotted region is the parameter space
allowed by neutron EDM upper bound in the plane of 
relative phase of $\l^{\prime}_{331}B^{\ast}_3$ and the
$|\l^{\prime}_{331}B^{\ast}_3|$ (left) and 
relative phase of $\l^{\prime}_{321}B^{\ast}_3$ and the
$|\l^{\prime}_{321}B^{\ast}_3|$ (right).
\label{scatter}}}

\end{document}